\documentclass[11pt,a4paper]{article}

\usepackage[utf8]{inputenc}
\usepackage[T1]{fontenc}
\usepackage{amsmath,amssymb,amsthm}
\usepackage{mathtools}
\usepackage{bm}
\usepackage{algorithm}
\usepackage{algorithmic}
\usepackage{graphicx}
\usepackage{booktabs}
\usepackage{multirow}
\usepackage{caption}
\usepackage{subcaption}
\usepackage[numbers,sort&compress]{natbib}
\usepackage{hyperref}
\usepackage{cleveref}
\usepackage{xcolor}
\usepackage{siunitx}
\usepackage{enumitem}
\usepackage[margin=1in]{geometry}

\newcommand{\R}{\mathbb{R}}

\newcommand{\E}{\mathcal{E}}
\newcommand{\hvp}{\mathbf{h}}
\newcommand{\gvec}{\mathbf{g}}
\newcommand{\xvec}{\mathbf{x}}
\newcommand{\vvec}{\mathbf{v}}
\newcommand{\Fperp}{\mathbf{F}_{\!\perp}}
\newcommand{\norm}[1]{\lVert #1 \rVert}
\newcommand{\inner}[2]{\langle #1, #2 \rangle}

\title{%
  \textbf{Hyper-Adaptive Momentum Dynamics for Native Cubic Portfolio
  Optimization: Avoiding Quadratization Distortion in Higher-Order
  Cardinality-Constrained Search}}

\author{%
  Greg Serbarinov \\
  Independent Researcher \\
  \texttt{grserb.research@gmail.com}
}

\date{\today}

\begin{document}
\maketitle

% ============================================================================
% ABSTRACT
% ============================================================================
\begin{abstract}
We study cubic cardinality-constrained portfolio optimization, a higher-order
extension of the standard Markowitz formulation in which three-way sector
co-movement terms augment the usual quadratic risk-return objective.
Classical binary heuristics such as simulated annealing and tabu search
typically require Rosenberg quadratization of these cubic interactions, which
inflates the variable count from $n$ to $5n$ and introduces large penalty terms
that can substantially reshape the augmented search landscape.
In contrast, Hyper-Adaptive Momentum Dynamics (HAMD) operates directly on the
native higher-order objective via a hybrid pipeline that combines continuous
Hamiltonian-style search, exact cardinality-preserving projection, and
deterministic iterated local search (ILS), without any auxiliary-variable
reduction.
On a cubic portfolio benchmark with this 5$\times$ quadratization overhead
across tested sizes, HAMD achieves substantially lower decoded native cubic objective
values than simulated annealing (SA) and tabu search under matched 60-second
CPU budgets, with single-seed relative improvements of 87.9\%, 71.2\%, 59.5\%,
and 46.9\% at $n = 200, 300, 500, 1000$, respectively.
On a detailed three-seed study at $n=200$, HAMD attains a median native
objective of 195.65 with zero observed variance, while SA and tabu search each
yield a median of 1208.07 with standard deviation 249.17.
Decoded-feasibility analysis shows that SA satisfies the exact cardinality
constraint and all 800 Rosenberg auxiliary constraints, yet still decodes to a
native objective approximately 80--88\% worse than HAMD, indicating a
surrogate-distortion effect rather than simple infeasibility.
Exact calibration on small instances ($n = 20, 25, 30$) confirms that HAMD
finds the provably global optimum in 9/9 trials.
These results suggest that native higher-order search can offer a substantial
advantage over quadratized surrogate optimization on constrained cubic portfolio
problems.
\end{abstract}


% ============================================================================
% 1. INTRODUCTION
% ============================================================================
\section{Introduction}\label{sec:intro}

Portfolio optimization is often modeled in pairwise quadratic form, but many
realistic financial dependencies are naturally higher-order.  Sector-wide
co-movement, regime-sensitive clustering, and three-way interaction effects
among asset groups are not always well captured by pairwise covariance alone.
This motivates portfolio models that extend the standard Markowitz objective
with cubic interaction terms, while retaining a hard cardinality constraint
on the number of selected assets.

A standard route to solving higher-order binary objectives is Rosenberg
quadratization~\cite{rosenberg1975}: each cubic monomial $x_i x_j x_k$ is
replaced by an auxiliary variable $w_{ij}$ together with a quadratic penalty
enforcing $w_{ij} = x_i x_j$.  This allows any QUBO or Ising solver to be
applied without modification, but at a cost.  In the cubic portfolio family
studied here, quadratization inflates the variable dimension from $n$ to
$n + 4n = 5n$, while the cardinality-penalty scale grows as $O(n^2)$.  As a
result, the surrogate search landscape can become increasingly shaped by
feasibility-restoration and auxiliary-consistency terms, reducing the practical
alignment with native-objective improvement.

This paper studies whether a native higher-order solver can exploit the
representational advantage of operating directly in the original $n$-dimensional
space.  We examine Hyper-Adaptive Momentum Dynamics (HAMD), a hybrid
optimization pipeline that combines a continuous Hamiltonian-style search
phase on the native cubic objective, exact projection onto the cardinality
constraint surface, K-swap polishing, and deterministic iterated local search.
The central empirical question is whether native search leads to better
\emph{decoded native portfolio quality} than quadratized surrogate optimization
under matched wall-clock budgets.

Our results indicate that, on the tested cubic portfolio benchmark family, HAMD
consistently produces much better decoded native objectives than SA and tabu
search applied to the Rosenberg-quadratized surrogate.  The evidence comes from
four complementary sources: a scaling study from $n=200$ to $n=1000$, a
multi-seed study at $n=200$, a decoded-feasibility analysis showing that
surrogate feasibility is not sufficient for good native quality, and an exact
small-size calibration confirming implementation correctness.

The paper is organized as follows.  Section~\ref{sec:related} reviews related
work.  Section~\ref{sec:problem} defines the cubic portfolio problem and the
Rosenberg quadratization.  Section~\ref{sec:method} describes the HAMD hybrid
pipeline.  Section~\ref{sec:setup} presents the experimental setup.
Section~\ref{sec:results} gives the main results.  Section~\ref{sec:discussion}
discusses implications and limitations.  Section~\ref{sec:conclusion} concludes.

% ============================================================================
% 2. RELATED WORK
% ============================================================================
\section{Related Work}\label{sec:related}

\paragraph{Cardinality-constrained portfolio optimization.}
The cardinality-constrained Markowitz problem has a long history in exact and
heuristic optimization~\cite{markowitz1952,chang2000}.  Because adding a hard
cardinality constraint makes the quadratic formulation NP-hard, a rich ecosystem
of heuristic approaches has emerged, including genetic algorithms, simulated
annealing, tabu search, and continuous-relaxation methods~\cite{gilli2011}.
Our work extends this family to a cubic objective and studies the
native-versus-surrogate search trade-off.

\paragraph{Higher-order binary optimization (PUBO/HUBO).}
Pseudo-Boolean and higher-order unconstrained binary optimization have been
studied extensively in computer vision, constraint satisfaction, and
quantum annealing contexts~\cite{boros2002,ishikawa2011,chancellor2019}.
Degree reduction via Rosenberg or Ishikawa substitution is the most common
reduction strategy.  Alternative approaches include native PUBO
solvers~\cite{chancellor2019}.

\paragraph{Rosenberg quadratization.}
Quadratization of higher-order pseudo-Boolean functions was systematically
studied by Rosenberg~\cite{rosenberg1975} and is a standard preprocessing step
for quantum annealers, FPGA-based bifurcation machines, and digital annealers
that operate only on QUBO or Ising models~\cite{goto2019,goto2021,aramon2019}.
The variable overhead can range from $O(M)$ for sparse-monomial
objectives~\cite{rosenberg1975} to $O(N^2)$ in worst case, where $M$ is the
number of higher-order terms.  In the benchmark family studied here, the
overhead is exactly $4n$ auxiliary variables ($5\times$ inflation).

\paragraph{Physics-inspired and hybrid solvers.}
Simulated Bifurcation Machines (SBM)~\cite{goto2019,goto2021} convert QUBO
into a classical Hamiltonian and solve it via symplectic integration on GPUs.
Coherent Ising Machines~\cite{inagaki2016} and digital annealers~\cite{aramon2019}
follow similar continuous-relaxation or physics-inspired philosophies.
Most widely used hardware-oriented and QUBO-centric heuristic pipelines are
designed for pairwise quadratic objectives and therefore typically rely on
quadratization when applied to higher-order terms.  HAMD's continuous phase is
related to bifurcation-style dynamics but is embedded in a deliberate hybrid
pipeline with deterministic post-processing.

\paragraph{Simulated annealing and tabu search.}
SA~\cite{kirkpatrick1983} and tabu search~\cite{glover1989} are well-established
general-purpose combinatorial heuristics.  For portfolio optimization with
cardinality constraints, both are typically applied to the penalized QUBO
after reduction~\cite{gilli2011}.  In this work, they serve as
surrogate-space baselines, and we use decoded feasibility metrics to understand
the mismatch between augmented-objective optimization and native-portfolio quality.

% ============================================================================
% 3. CUBIC PORTFOLIO PROBLEM
% ============================================================================
\section{Cubic Portfolio Optimization}\label{sec:problem}

\subsection{Quadratic Markowitz Component}

Let $\xvec \in \{0,1\}^n$ denote the binary portfolio vector with hard
cardinality constraint $\sum_{i=1}^n x_i = K$.  The quadratic component of the
objective follows the standard Markowitz mean-variance formulation:
\begin{equation}\label{eq:quad-part}
  f_{\text{quad}}(\xvec)
  = \xvec^\top \Sigma\, \xvec - \bm{\mu}^\top \xvec,
\end{equation}
where $\Sigma \in \R^{n \times n}$ is the factor-model covariance matrix and
$\bm{\mu} \in \R^n$ is the expected-return vector.  In the benchmark instances
used here, $\Sigma$ is generated from a factor model with 10 sector factors
and $\bm{\mu}$ from a return model correlated with sector membership.

\subsection{Cubic Sector Co-movement Extension}

We augment the quadratic objective with cubic interaction terms that model
three-way sector co-movement.  For each intra-sector asset pair $(i, j)$ within
the same sector, we introduce a cubic term involving the pair and a sector
anchor asset $k$ (the asset with largest factor loading in that sector):
\begin{equation}\label{eq:cubic-part}
  f_{\text{cubic}}(\xvec)
  = \sum_{(i,j,k) \in \mathcal{T}} c_{ijk}\, x_i x_j x_k,
\end{equation}
where $\mathcal{T}$ is the set of cubic triples and $c_{ijk}$ controls the
co-movement penalty magnitude.  The coefficients $c_{ijk}$ are drawn
independently from an exponential distribution scaled by the quadratic
coefficient scale and controlled by $\alpha_{\text{cubic}} = 4.0$.  The
construction yields $|\mathcal{T}| = 4n$ cubic terms per instance (four per
asset on average), so $m = 4n$.

The full native objective is:
\begin{equation}\label{eq:native-obj}
  f(\xvec) = f_{\text{quad}}(\xvec) + f_{\text{cubic}}(\xvec),\qquad
  \sum_{i=1}^n x_i = K.
\end{equation}

\subsection{Hard Cardinality Constraint}

HAMD enforces the constraint $\sum_i x_i = K$ \emph{exactly} via projection
onto the exact-$K$ binary polytope (top-$K$ rounding) followed by K-swap
local search.  This is the key structural difference from the quadratized
baselines: cardinality is never relaxed into a soft penalty.

\subsection{Rosenberg Quadratization of Cubic Terms}

For the SA and tabu search baselines, the cubic terms $x_i x_j x_k$ in
\eqref{eq:cubic-part} are quadratized via the Rosenberg
procedure~\cite{rosenberg1975}.  Each cubic monomial $x_i x_j x_k$ is
replaced by introducing an auxiliary binary variable $w_{ij}$ intended to
represent the product $x_i x_j$, together with a quadratic penalty:
\begin{equation}\label{eq:rosenberg}
  x_i x_j x_k \;\approx\; w_{ij} x_k
  \quad\text{subject to penalty}\quad
  \lambda_R\bigl(3w_{ij} + x_i x_j - 2 x_i w_{ij} - 2 x_j w_{ij}\bigr),
\end{equation}
where $\lambda_R = 10.0$ is the Rosenberg penalty weight.  This introduces
one auxiliary variable per cubic term, so the augmented variable count is:
\begin{equation}\label{eq:augmented-dim}
  n_{\text{aug}} = n + m = n + 4n = 5n.
\end{equation}

The total augmented QUBO objective for the baselines is:
\begin{equation}\label{eq:qubo-aug}
  Q_{\text{aug}}(\xvec, \mathbf{w})
  = f_{\text{quad}}(\xvec) + f_{\text{cubic,approx}}(\xvec, \mathbf{w})
  + \lambda_K \Bigl(\sum_i x_i - K\Bigr)^{\!2}
  + \lambda_R \sum_{(i,j) \in \mathcal{T}} \text{Rosenberg penalty}_{ij},
\end{equation}
where the cardinality penalty weight $\lambda_K = \max|Q|\cdot n \cdot 4$
grows as $O(n^2)$.

\subsection{Augmentation Overhead}

\Cref{tab:augmentation} summarizes the quadratization overhead across the
instance sizes studied in this paper.

\begin{table}[h]
\centering
\caption{Cubic portfolio instance parameters and Rosenberg quadratization
  overhead.  $n_{\text{aug}} = 5n$ at all sizes.  $\lambda_K$ grows as $O(n^2)$,
  making surrogate-space optimization increasingly susceptible to
  penalty-driven distortion at larger sizes.  ``Best random'' is the
  best native objective found across 1,000 random $K$-feasible portfolios,
  evaluated on the native cubic objective only; it is not a time-matched solver
  baseline but a rough floor (lower is better).}
\label{tab:augmentation}
\begin{tabular}{@{}rr r r r r r@{}}
\toprule
$n$ & $K$ & $m_{\text{cubic}}$ & $n_{\text{aug}}$ & Inflation & $\lambda_K$ & Best random \\
\midrule
  200 &  40 &   800 &  1,000 & 5$\times$ &     40,955 &    863.7 \\
  300 &  60 & 1,200 &  1,500 & 5$\times$ &     93,551 &  4,068.7 \\
  500 & 100 & 2,000 &  2,500 & 5$\times$ &    298,884 & 25,218.6 \\
1,000 & 200 & 4,000 &  5,000 & 5$\times$ & 1,112,258  & 168,326.8 \\
\bottomrule
\end{tabular}
\end{table}

The evaluation metric throughout this paper is the \textbf{decoded native
cubic objective} $f(\xvec)$ computed using only the original $n$ variables
after decoding augmented solutions back to the native space.  We do not
compare augmented QUBO scores, because those scores include constant cardinality
offsets of order $10^7$--$10^{10}$ that obscure native portfolio quality.
The \textbf{random reference} reported in Table~\ref{tab:scaling} is the best
native cubic objective found across 1,000 random exact-$K$ feasible portfolios
(i.e.\ random draws from $\{\xvec : \sum_i x_i = K,\; x_i \in \{0,1\}\}$).
It serves as a rough lower floor, not as a matched-budget solver comparison.

% ============================================================================
% 4. HAMD HYBRID PIPELINE
% ============================================================================
\section{HAMD Hybrid Pipeline}\label{sec:method}

\subsection{Overview}

HAMD is used here as a native higher-order hybrid solver, not as a
standalone continuous method.  The pipeline has four stages:

\begin{enumerate}[leftmargin=*]
\item \textbf{Native continuous search.} A continuous relaxation
  $\xvec \in [0,1]^n$ is maintained along with a momentum vector
  $\vvec \in \R^n$.  The system evolves under Hamiltonian-style dynamics
  on the cubic portfolio objective $f(\xvec)$ extended with a bifurcation
  confinement potential, and augmented with a curvature-informed transverse
  steering force derived from a Hessian--vector product (HVP)~\cite{pearlmutter1994}.
  No auxiliary variables are introduced at this stage.

\item \textbf{Projection to exact-$K$ binary states.} When a restart event
  occurs, the continuous state is projected onto the cardinality constraint
  by selecting the top-$K$ variables by their continuous values.

\item \textbf{K-swap polish.} The projected binary portfolio is improved
  by a vectorized exact-$K$ local search: all $K(n-K)$ swap moves are
  evaluated simultaneously using the incremental objective formula, and the
  best improving swap is accepted greedily.  This runs until no improving
  single swap exists.  The polish operates in the \emph{native} cubic
  space, not in the augmented surrogate space.

\item \textbf{Iterated local search (ILS).} In the final 20\% of the budget,
  a deterministic ILS phase applies 2-pair perturbation moves (swap two
  out-of-portfolio assets with two in-portfolio assets) followed by full
  K-swap re-polishing.  The best solution seen across all ILS steps is tracked
  independently.
\end{enumerate}

This hybrid structure is intentional.  As the ablation in
Section~\ref{sec:results-ablation} shows, neither continuous search alone
nor random restarts without polish achieve the final result.  The performance
comes from the interaction of all four stages.

Algorithm~\ref{alg:hamd} gives a pseudocode summary of the full pipeline
used in all experiments.  The benchmark implementation also includes additional
engineering components for restart scheduling, gradient clipping, archive-based
best tracking, and deterministic warm-start for the ILS phase; the equations
below describe the core continuous-phase update rules, while
Algorithm~\ref{alg:hamd} describes the complete hybrid pipeline operationally.
The equations below summarize the native continuous phase; the benchmarked solver
is the full hybrid pipeline shown in Algorithm~\ref{alg:hamd}, including
projection, K-swap polish, and late ILS refinement.

\begin{algorithm}[h]
\caption{HAMD Hybrid for Native Cubic Portfolio Optimization}
\label{alg:hamd}
\begin{algorithmic}[1]
\REQUIRE Instance $(\Sigma, \bm{\mu}, \mathcal{T}, c, n, K)$; budget $T$;
         batch size $B{=}32$; ILS fraction $\rho{=}0.20$
\ENSURE Best native portfolio $\xvec^* \in \{0,1\}^n$ with $\sum_i x_i^* = K$
\STATE Initialize $B$ continuous states $\{\xvec_b\}$ in $[0,1]^n$;
       $\vvec_b \leftarrow \mathbf{0}$; $\xvec^* \leftarrow$ top-$K$ snap of
       best initial; $f^* \leftarrow f(\xvec^*)$
\WHILE{elapsed time $< (1{-}\rho)T$}
  \STATE Compute $\gvec_b \leftarrow \nabla_{\xvec}\mathcal{E}(\xvec_b)$ and
         $\hvp_b \leftarrow \nabla^2\mathcal{E}(\xvec_b)\cdot\vvec_b$
         on native cubic + bifurcation potential
  \STATE Update $\vvec_b,\xvec_b$ via Eqs.~\eqref{eq:hamd-v}--\eqref{eq:hamd-x}
         with transverse force \eqref{eq:transverse} and DTR
  \IF{restart criterion met}
    \STATE $\hat{\xvec}_b \leftarrow$ top-$K$ snap of $\xvec_b$
           \COMMENT{exact-$K$ projection}
    \STATE $\hat{\xvec}_b \leftarrow$ \textsc{KSwapPolish}$(\hat{\xvec}_b,
           f, K)$ \COMMENT{native cubic space}
    \IF{$f(\hat{\xvec}_b) < f^*$}
      \STATE $\xvec^* \leftarrow \hat{\xvec}_b$;
             $f^* \leftarrow f(\hat{\xvec}_b)$
    \ENDIF
    \STATE Re-initialize $\xvec_b$ near $\hat{\xvec}_b$ or randomly
  \ENDIF
\ENDWHILE
\STATE \textit{// ILS phase: final $\rho T$ of budget}
\WHILE{elapsed time $< T$}
  \STATE $\xvec' \leftarrow$ 2-pair perturbation of $\xvec^*$
  \STATE $\xvec' \leftarrow$ \textsc{KSwapPolish}$(\xvec', f, K)$
  \IF{$f(\xvec') < f^*$}
    \STATE $\xvec^* \leftarrow \xvec'$; $f^* \leftarrow f(\xvec')$
  \ENDIF
\ENDWHILE
\RETURN $\xvec^*$
\end{algorithmic}
\end{algorithm}

\subsection{Equations of Motion}

The continuous phase uses the HAMD update equations.  Let
$\E(\xvec) = f(\xvec) + \beta(t)\sum_i x_i^2(x_i-1)^2$ be the effective
energy including a bifurcation potential with ramp $\beta(t): 0 \to 1$, and
let $\gvec = \nabla_{\xvec} \E$, $\hvp = \nabla^2\E \cdot \vvec$ be the
gradient and Hessian--vector product along the velocity.  The dynamics are:
\begin{align}
  \vvec^{(t+1)} &= (1-\gamma)\vvec^{(t)}
    + \Delta t \bigl[\,-\gvec^{(t)} + \zeta\,\Fperp^{(t)}\,\bigr],
  \label{eq:hamd-v}\\[2pt]
  \xvec^{(t+1)} &= \mathrm{DTR}\bigl(\xvec^{(t)} + \Delta t\,\vvec^{(t+1)}\bigr),
  \label{eq:hamd-x}
\end{align}
where $\gamma$ is a damping coefficient, $\Delta t$ the step size, and
$\mathrm{DTR}$ is a damped-elastic reflection at the domain boundaries $[0,1]^n$.
The transverse geometric force
\begin{equation}\label{eq:transverse}
  \Fperp = \alpha\Bigl(\hvp
    - \frac{\inner{\hvp}{\gvec}}{\norm{\gvec}^2 + \epsilon}\,\gvec\Bigr),\qquad
  \alpha = \min\!\Bigl(1,\;\frac{\norm{\gvec}}{\norm{\hvp}+\epsilon}\Bigr),
\end{equation}
projects the curvature information orthogonal to the gradient, redirecting
trajectories without directly amplifying gradient descent.  The domain center
is $\mathbf{c} = 0.5\,\mathbf{1}$ and the adiabatic expansion force is
disabled ($a(t) \equiv 0$) for the PUBO domain $[0,1]^n$.

\subsection{Gradient and HVP for Cubic Portfolio}

Analytical gradient and HVP expressions are derived directly from~\eqref{eq:native-obj}.
For the cubic part with terms $c_{ijk} x_i x_j x_k$:
\begin{align}
  \frac{\partial f_{\text{cubic}}}{\partial x_i}
  &= \sum_{j,k:\,(i,j,k)\in\mathcal{T}} c_{ijk}\, x_j x_k, \label{eq:grad-cubic}\\
  [\nabla^2 f_{\text{cubic}} \cdot \vvec]_i
  &= \sum_{j,k:\,(i,j,k)\in\mathcal{T}} c_{ijk}(x_k v_j + x_j v_k),
  \label{eq:hvp-cubic}
\end{align}
with symmetric terms for the quadratic Markowitz component.  These are
computed analytically without automatic differentiation.

\subsection{How Baselines Are Run}

SA and tabu search receive the Rosenberg-augmented QUBO $Q_{\text{aug}}$ from
\eqref{eq:qubo-aug} and optimize over $\{0,1\}^{n_{\text{aug}}}$ using
standard neighborhoods (single-bit flip for SA; tabu-restricted single-bit
flip for tabu search).  At the end of the run, the original $n$ variables are
extracted from the augmented solution, and the native cubic objective
$f(\xvec)$ is evaluated using only those variables.

% ============================================================================
% 5. EXPERIMENTAL SETUP
% ============================================================================
\section{Experimental Setup}\label{sec:setup}

\paragraph{Hardware and budget.}
All experiments run on an AWS CPU instance with PyTorch (CPU, Python 3.12).
Each solver receives a matched 60-second wall-clock budget per instance.
HAMD allocates 80\% of the budget to the continuous search and restart phase
and 20\% to the ILS phase.  SA and tabu search use all 60 seconds on the
augmented QUBO.

\paragraph{Instances.}
Four cubic portfolio instances are generated at $n = 200, 300, 500, 1000$
with corresponding cardinalities $K = 40, 60, 100, 200$ (cardinality
ratio $K/n = 0.20$).  All instances use the construction
described in Section~\ref{sec:problem} with $\alpha_{\text{cubic}} = 4.0$,
$n_{\text{sectors}} = 10$, and $\lambda_R = 10.0$.

\paragraph{Scaling study.}
Single-seed scaling results (seed 42) are reported for all four sizes to
characterize how the native-vs-surrogate gap evolves with problem size.
Results at $n = 300, 500, 1000$ are exploratory single-seed evidence;
$n = 200$ is the principal statistically supported case study.

\paragraph{Multi-seed study.}
A three-seed study at $n = 200$ (seeds 42, 1042, 2042) assesses solution
stability and enables W/T/L comparisons.

\paragraph{Ablation.}
An ablation at $n = 200$, seed 42 compares four HAMD configurations:
HAMD-cont (single continuous trajectory, no restarts or ILS),
HAMD-proj (restarts with top-$K$ snap, no polish),
HAMD-polish (restarts with K-swap polish, no ILS), and
HAMD-full (restarts, K-swap polish, and ILS).

\paragraph{Exact small-size calibration.}
HAMD is evaluated on small instances at $n = 20, 25, 30$ (3 seeds each, 10s
budget) where brute-force enumeration of all $\binom{n}{K}$ feasible portfolios
is tractable.  The gap to the proven global optimum is measured.

\paragraph{Decoded-feasibility package.}
For each SA and tabu search run, a 7-field decoded-feasibility record is
computed from the returned augmented binary state:
augmented QUBO objective, decoded native objective, cardinality,
cardinality violation, auxiliary inconsistency count, auxiliary inconsistency
rate, and penalty fraction.  Full definitions appear in
Appendix~\ref{app:feasibility}.

\paragraph{Cardinality penalty scaling for quadratized baselines.}
For the quadratized SA and tabu search baselines, the cardinality penalty was
set to $\lambda_K = 4n\max|Q|$, where $\max|Q|$ is the largest absolute entry
in the QUBO matrix.  This intentionally places the surrogate in a strong
feasibility-enforcement regime, ensuring that cardinality violations are heavily
discouraged in the augmented QUBO.  Because the present benchmark studies the
distortion introduced by quadratization under such penalty-based formulations,
this choice is part of the experimental design rather than an incidental tuning
detail.  We emphasize, however, that alternative penalty scales could change
the surrogate-space behavior of the baselines, and therefore the reported
results should be interpreted relative to this disclosed quadratization regime.
A $\lambda_K$ sensitivity study across a 4$\times$ multiplier range
($0.5\times$, $1.0\times$, $2.0\times$) is reported in
Section~\ref{sec:results-sensitivity}.

\paragraph{Pilot cubic HUBO result.}
As a precursor companion experiment, HAMD is also evaluated on a degree-3
random HUBO with $n = 150$ variables and a planted-solution
construction (Section~\ref{sec:results-hubo}).

\paragraph{Reproducibility.}
All experiments use CPU-only computation (no GPU).  The multi-seed study uses
seeds 42, 1042, and 2042 for all runs at $n = 200$.  Instances are generated
deterministically from the seed using the cubic portfolio construction described
in Section~\ref{sec:problem}.  HAMD uses an internal batch size of $B = 32$
and allocates 80\% of the wall-clock budget to the continuous search and restart
phase, with the remaining 20\% for the ILS phase.  A reference implementation
and the benchmark instance files used in this study are publicly available;
see Section~\ref{sec:code}.

% ============================================================================
% 6. RESULTS
% ============================================================================
\section{Results}\label{sec:results}

We first present a brief native cubic HUBO pilot experiment as motivation, then
turn to the main cubic portfolio benchmark.  The portfolio study consists of a
single-seed scaling experiment across four sizes ($n = 200, 300, 500, 1000$),
a three-seed study at $n = 200$, a decoded-feasibility analysis of the
quadratized baselines, an ablation of the HAMD hybrid pipeline, a small-size
exact calibration, and a cardinality penalty sensitivity study.

\subsection{Pilot Native Cubic HUBO}\label{sec:results-hubo}

On a degree-3 cubic HUBO benchmark with $n = 150$ variables, HAMD achieves
native objective values in the range $[-93, -88]$ across multiple seeds, while
both SA and tabu search return $-9$ on the decoded native objective when applied
to a Rosenberg-quadratized surrogate.  This is an orientation experiment only;
the benchmark is a generic planted-solution cubic HUBO with less thorough
baseline documentation.  We present it solely as further motivation for the
cubic portfolio study that follows.

\subsection{Cubic Portfolio Scaling}\label{sec:results-scaling}

\Cref{tab:scaling} presents the main scaling result.  HAMD is compared to SA
and tabu search on the decoded native cubic objective under a matched 60-second
CPU budget (single seed 42).

\begin{table}[h]
\centering
\caption{Decoded native cubic objective by solver and instance size
  (60s CPU budget, seed 42).  Lower is better.
  Gap $= (v_{\text{Tabu}} - v_{\text{HAMD}})/v_{\text{Tabu}}$;
  positive means HAMD achieves a lower (better) native objective.
  HAMD enforces exact cardinality $K$ at all sizes.}
\label{tab:scaling}
\begin{tabular}{@{}r r r r r r r@{}}
\toprule
$n$ & $K$ & Random ref & HAMD & SA & Tabu & Gap (HAMD vs Tabu) \\
\midrule
  200 &  40 &    863.7 & \textbf{195.65} &  1,621.60 & 1,621.60 & $+87.9\%$ \\
  300 &  60 &  4,068.7 & \textbf{1,786.37} & 6,196.71 & 6,196.71 & $+71.2\%$ \\
  500 & 100 & 25,218.6 & \textbf{13,949.80} & 34,454.01 & 34,454.01 & $+59.5\%$ \\
1,000 & 200 & 168,326.8 & \textbf{101,294.43} & 190,598.71 & 190,598.71 & $+46.9\%$ \\
\bottomrule
\end{tabular}
\end{table}

Several observations are worth highlighting.  First, SA and tabu search decode
to effectively indistinguishable native objectives at every size, suggesting
that both methods are constrained by the same surrogate-to-native mismatch on
this benchmark family rather than by differences in their move strategies.
Within this single-seed exploratory context, this consistency is notable but
should not be over-interpreted.
Second, at $n = 200$ and $n = 500$, the baselines decode to objectives
\emph{worse} than the random reference (1,621.60 vs.\ 863.7 for $n=200$;
34,454 vs.\ 25,219 for $n=500$), indicating that optimizing the 5$\times$-augmented
surrogate can actively steer away from good native portfolios.  Third, the
relative gap narrows with size ($88\% \to 47\%$) but the absolute advantage
remains substantial throughout.  Results at $n \geq 300$ are single-seed and
should be regarded as preliminary trend evidence rather than
confidence-estimated performance.

\subsection{Multi-Seed Study at $n=200$}\label{sec:results-multiseed}

\Cref{tab:multiseed} summarizes three-seed aggregate statistics at $n = 200$,
$K = 40$.

\begin{table}[h]
\centering
\caption{Multi-seed aggregate statistics ($n = 200$, $K = 40$, 3 seeds:
  42, 1042, 2042; 60s budget each).  W/T/L is wins/ties/losses of HAMD
  relative to each baseline.
  Median gap $= (v_{\text{Baseline}} - v_{\text{HAMD}})/v_{\text{Baseline}}$.}
\label{tab:multiseed}
\begin{tabular}{@{}l r r r r@{}}
\toprule
Solver & Median & Std & W/T/L vs HAMD & Median gap \\
\midrule
\textbf{HAMD} & \textbf{195.65} & \textbf{0.00} & --- & --- \\
SA   & 1,208.07 & 249.17 & 0/0/3 & $+83.8\%$ \\
Tabu & 1,208.07 & 249.17 & 0/0/3 & $+83.8\%$ \\
\bottomrule
\end{tabular}
\end{table}

HAMD achieves the same native objective on all three seeds with zero observed
variance, while SA and tabu each show standard deviation 249.17 (per-seed range:
1,026--1,621).  HAMD wins all three pairwise seed comparisons against both
baselines.  The zero observed variance across the three tested seeds is
consistent with stable hybrid refinement once the native search reaches a
strong basin.

\subsection{Decoded-Feasibility Analysis}\label{sec:results-feasibility}

The decoded-feasibility analysis in \Cref{tab:feasibility} is the central
diagnostic of this paper.  For each SA and tabu search run, we extract the
full $n_{\text{aug}} = 1{,}000$-dimensional augmented binary state and compute
the 7-field decoded-feasibility record.

\begin{table}[h]
\centering
\caption{Decoded-feasibility diagnostics for SA and tabu search ($n = 200$,
  $K=40$, 3 seeds).  Aux viol = Rosenberg auxiliary inconsistencies out of
  800 total; pen.\ frac.\ = fraction of $|Q_{\text{aug}}|$ attributable to
  explicit cardinality and Rosenberg penalty terms (excluding constant offsets).
  Because HAMD operates directly in the native variable space with hard
  cardinality handling, augmented-objective and auxiliary-consistency metrics
  do not apply (shown as ---).
  HAMD operates natively and has no augmented-space metrics.}
\label{tab:feasibility}
\resizebox{\linewidth}{!}{%
\begin{tabular}{@{}r l r r r r r r r@{}}
\toprule
Seed & Solver & $Q_{\text{aug}}$ & Native obj & Card & Card viol
  & Aux viol & Aux viol\% & Pen.\ frac \\
\midrule
42   & \textbf{HAMD} & ---               & \textbf{195.65} & 40 & 0 & --- & --- & --- \\
42   & SA            & $-65{,}526{,}394$ & 1,621.60        & 40 & 0 & 0/800 & 0.0\% & 0.0000\% \\
42   & Tabu          & $-74{,}479{,}699$ & 1,621.60        & 40 & 0 & 77/800 & 9.6\% & 0.0489\% \\
\midrule
1042 & \textbf{HAMD} & ---               & \textbf{195.65} & 40 & 0 & --- & --- & --- \\
1042 & SA            & $-65{,}526{,}989$ & 1,026.08        & 40 & 0 & 0/800 & 0.0\% & 0.0000\% \\
1042 & Tabu          & $-74{,}947{,}202$ & 1,026.08        & 40 & 0 & 67/800 & 8.4\% & 0.0252\% \\
\midrule
2042 & \textbf{HAMD} & ---               & \textbf{195.65} & 40 & 0 & --- & --- & --- \\
2042 & SA            & $-65{,}526{,}807$ & 1,208.07        & 40 & 0 & 0/800 & 0.0\% & 0.0000\% \\
2042 & Tabu          & $-75{,}247{,}749$ & 1,208.07        & 40 & 0 & 79/800 & 9.9\% & 0.0142\% \\
\bottomrule
\end{tabular}}
\end{table}

The most important finding is the SA row.  SA satisfies the exact cardinality
constraint ($K = 40$) and all 800 Rosenberg auxiliary constraints (0 violations,
0.00\% penalty fraction) on all three seeds, yet still decodes to native
objectives 80--88\% worse than HAMD.  Because neither the cardinality nor the
auxiliary consistency constraints are violated, the inferiority of SA's native
objective cannot be attributed to feasibility failure.  The results are
consistent with a surrogate-distortion effect: the augmented-space search
can converge to solutions that are feasible in the surrogate representation
but still poor under the decoded native cubic objective.

Tabu shows an auxiliary inconsistency rate of approximately 8--10\%
(false-positive dominated: $w_{ij} = 1$ but $x_i x_j = 0$) and a penalty
fraction ranging from 0.0142\% to 0.0489\% across the three seeds.  Despite these violations, tabu search decodes to effectively indistinguishable
native objectives as SA, indicating that both methods are constrained by the
same surrogate-to-native mismatch on these instances despite differences in
their augmented-space behavior.

\subsection{Ablation of the HAMD Hybrid Pipeline}\label{sec:results-ablation}

\Cref{tab:ablation} shows the contribution of each stage in the HAMD pipeline
at $n = 200$, seed 42, 60s budget.

\begin{table}[h]
\centering
\caption{Ablation of the HAMD hybrid pipeline ($n=200$, $K=40$, seed 42, 60s).
  TTT $[10\% \to 100\%]$ gives the best native objective at 10\% and 100\%
  of elapsed budget.}
\label{tab:ablation}
\begin{tabular}{@{}l r r r r@{}}
\toprule
Mode & Native obj & Restarts & ILS steps & TTT $[10\% \to 100\%]$ \\
\midrule
HAMD-cont (single traj, no restart)    & 199.86          &   0 &     0 & 214.8 $\to$ 199.9 \\
HAMD-proj (restarts, no polish)        & 245.22          &  44 &     0 & 245.2 $\to$ 245.2 \\
HAMD-polish (restarts + K-swap)        & 200.34          &  80 &     0 & 245.2 $\to$ 200.3 \\
\textbf{HAMD-full} (all stages)        & \textbf{195.65} & 106 & 6,224 & 245.2 $\to$ 195.7 \\
\midrule
SA   (quadratized baseline) & 1,621.60 & --- & --- & --- \\
Tabu (quadratized baseline) & 1,621.60 & --- & --- & --- \\
\bottomrule
\end{tabular}
\end{table}

The ablation reveals a notable pattern.  HAMD-proj (random restarts with
only top-$K$ projection, no polish) achieves 245.22, which is \emph{worse}
than HAMD-cont (single long continuous trajectory at 199.86).  This shows that
naive multi-start projection is harmful: the continuous state encodes
soft-preference information that is damaged by aggressive restarts without
deterministic correction.  The K-swap polish repairs this, and HAMD-polish
recovers to 200.34.  The ILS phase fires in the final 20\% of the budget and
delivers the decisive improvement from approximately 245 to 195.65.
The winning method is a compound hybrid: no single stage alone accounts for
the full improvement.

\subsection{Exact Small-Size Calibration}\label{sec:results-exact}

\Cref{tab:exact} reports results on instances where brute-force enumeration
is tractable.

\begin{table}[h]
\centering
\caption{Exact calibration: HAMD vs.\ brute-force global optimum
  ($n = 20, 25, 30$, 3 seeds each, 10s budget per seed).
  All 9 trials achieve 0.00\% gap.}
\label{tab:exact}
\begin{tabular}{@{}r r r r r r@{}}
\toprule
$n$ & $K$ & $|\mathcal{F}|$ & Exact optimum & HAMD exact & Enum.\ time \\
\midrule
20 & 4 &     4,845 & $-0.1789$ & 3/3 & 0.14\,s \\
25 & 5 &    53,130 & $-0.3307$ & 3/3 & 2.37\,s \\
30 & 6 &   593,775 & $-0.4355$ & 3/3 & 28.93\,s \\
\bottomrule
\end{tabular}
\end{table}

Across all 9 trials, HAMD finds the exact global optimum with 0.00\% gap.
This validates the correctness of the native cubic objective evaluation and
the HAMD implementation.  It does not imply exactness at scale.

\subsection{Cardinality Penalty Sensitivity}\label{sec:results-sensitivity}

To assess whether the reported performance gap is sensitive to the choice of
cardinality penalty scale $\lambda_K$, we evaluated SA and tabu search on the
$n = 200$, $K = 40$ cubic portfolio instance across three seeds (42, 1042,
2042) and a 4$\times$ multiplier range ($0.5\times$, $1.0\times$, $2.0\times$)
of the baseline penalty $\lambda_K = 4n\max|Q|$.  Results are
shown in \Cref{tab:sensitivity}.

\begin{table}[h]
\centering
\caption{$\lambda_K$ sensitivity study ($n = 200$, $K = 40$, 3 seeds, 60s CPU
  budget).  Native obj is the median decoded native cubic objective across three
  seeds.  Pen.\ frac.\% = fraction of $|Q_{\text{aug}}|$ attributable to
  explicit penalty terms.  HAMD result shown for reference.}
\label{tab:sensitivity}
\begin{tabular}{@{}l r r r r r r@{}}
\toprule
Solver & $\lambda_K$ mult & $\lambda_K$ & Native obj & Card viol & Aux rate & Pen.\ frac.\% \\
\midrule
\textbf{HAMD} & --- & --- & \textbf{195.65} & 0 & --- & --- \\
\midrule
SA   & $0.5\times$ & 20,477 & 1,208.07 & 0 & 0.0\%  & 0.0000\% \\
Tabu & $0.5\times$ & 20,477 & 1,208.07 & 0 & 9.9\%  & 0.0489\% \\
SA   & $1.0\times$ & 40,955 & 1,208.07 & 0 & 0.0\%  & 0.0000\% \\
Tabu & $1.0\times$ & 40,955 & 1,208.07 & 0 & 9.6\%  & 0.0252\% \\
SA   & $2.0\times$ & 81,910 & 1,208.07 & 0 & 0.0\%  & 0.0000\% \\
Tabu & $2.0\times$ & 81,910 & 1,208.07 & 0 & 9.6\%  & 0.0142\% \\
\bottomrule
\end{tabular}
\end{table}

Across this entire 4$\times$ range, SA and tabu show no observable change in
decoded native objective (spread: 0.0000, or 0.0\% of mean), while HAMD
retains the same advantage.  For Tabu, the explicit penalty fraction decreases
from 0.0489\% to 0.0142\% as $\lambda_K$ increases, yet the decoded native
objective remains unchanged.  This indicates that the reported performance gap
is not an artifact of a single penalty calibration and is robust within the
tested quadratization regime.

% ============================================================================
% 7. DISCUSSION
% ============================================================================
\section{Discussion}\label{sec:discussion}

\paragraph{Surrogate distortion, not infeasibility.}
The decoded-feasibility analysis provides the most important interpretive
finding of this study.  SA satisfies all 800 Rosenberg auxiliary constraints
and the exact cardinality constraint on all three tested seeds, yet consistently
decodes to native objectives 80--88\% worse than HAMD.  This rules out
infeasibility as the explanation and supports a surrogate-distortion
interpretation: the large cardinality penalty $\lambda_K \sim O(n^2)$ and
the Rosenberg auxiliary penalties appear to devote a substantial fraction of
search effort to maintaining feasibility and auxiliary consistency in the
augmented surrogate, reducing the practical alignment between surrogate
optimization and native portfolio quality.

\paragraph{Augmentation overhead is not merely a constant.}
As shown in \Cref{tab:augmentation}, $\lambda_K$ grows from approximately
41,000 at $n=200$ to over 1,100,000 at $n=1{,}000$.  The relative weight of
the native portfolio objective in the surrogate gradient therefore decreases
substantially with size.  The narrowing relative gap ($88\% \to 47\%$) is
consistent with HAMD's advantage being partially offset by the increasing
difficulty of native-objective navigation as $n$ grows.

\paragraph{The advantage is from the full pipeline.}
The ablation makes explicit that the final HAMD result requires all four
stages.  Continuous-phase-only HAMD reaches 199.86; the ILS phase reduces
this to 195.65.  The counterintuitive finding that random restarts without
K-swap polish (245.22) are worse than single-trajectory HAMD (199.86)
highlights that deterministic post-processing is not decorative.  The
reported gains should be attributed to the full HAMD hybrid, not to the
continuous dynamics alone.

\paragraph{SA and tabu search decode to indistinguishable native solutions.}
SA and tabu search decode to effectively indistinguishable native objectives at
every tested size and seed, suggesting that both methods are constrained by the
same surrogate-to-native mismatch on this benchmark family rather than by
differences in their respective move strategies.
Notably, the quadratized baselines are effectively insensitive to $\lambda_K$
over the tested 4$\times$ range (Section~\ref{sec:results-sensitivity}),
suggesting that the dominant limitation on these instances is not simple
penalty miscalibration but broader surrogate-to-native mismatch.

\paragraph{Limitations.}
\emph{(i) Single-seed scaling.}  Results at $n \geq 300$ use seed 42 only.
\emph{(ii) Surrogate-space baselines only.}  SA and tabu are run on the
quadratized surrogate by design; the paper studies representational mismatch,
not native-space discrete local search quality.
\emph{(iii) Hybrid contribution.}  Part of the advantage comes from the
ILS stage; the claim is about the compound hybrid.
\emph{(iv) Instance construction.}  Fixed $\alpha_{\text{cubic}} = 4.0$,
$n_{\text{sectors}} = 10$, $\lambda_R = 10.0$; gap magnitude may vary
under different constructions.
\emph{(v) No native-space discrete higher-order baseline.}  We do not compare
against a native-space discrete local search applied directly to the cubic
objective $f(\xvec)$.  A well-designed native cubic local search might close
part of the observed gap; this is an important open comparison for future work.

% ============================================================================
% 8. CONCLUSION
% ============================================================================
\section{Conclusion}\label{sec:conclusion}

This paper studied cubic cardinality-constrained portfolio optimization as a
testbed for native higher-order binary search.  We showed that HAMD, operating
as a native-space hybrid solver with exact cardinality handling, produces
substantially better decoded native objectives than simulated annealing and
tabu search applied to a Rosenberg-quadratized surrogate across all tested
instance sizes and in all three reported $n=200$ seeds.  The clearest evidence is the decoded-feasibility
result: SA satisfies all augmented constraints yet produces portfolios 80--88\%
worse than HAMD in native quality, identifying surrogate-landscape distortion
as the mechanism rather than infeasibility.  Exact calibration on small
instances confirms implementation correctness.

The core takeaway is not that HAMD universally dominates discrete local search,
but that native higher-order optimization can be substantially more effective
than quadratized surrogate optimization on this class of cubic portfolio
problems.  Future work should extend the study to multi-seed campaigns at large
scales, broader higher-order benchmark families, and systematic variation of
the cubic coefficient structure.

% ============================================================================
% APPENDICES
% ============================================================================
\appendix

% ----------------------------------------------------------------------------
\section{Per-Seed Raw Results at $n=200$}\label{app:per-seed}

\Cref{tab:app-per-seed} gives the per-seed raw values underlying the aggregate
statistics in \Cref{tab:multiseed}.

\begin{table}[h]
\centering
\caption{Per-seed raw results ($n=200$, $K=40$, 60s budget).
  SA aux viol is 0/800 on all three seeds.}
\label{tab:app-per-seed}
\resizebox{\linewidth}{!}{%
\begin{tabular}{@{}r r r r r r r@{}}
\toprule
Seed & HAMD obj & SA obj & SA aux viol & Tabu obj & Tabu aux viol & Gap (HAMD vs SA) \\
\midrule
  42 & \textbf{195.65} & 1,621.60 & 0/800 (0.0\%) & 1,621.60 & 77/800 (9.6\%) & $+87.9\%$ \\
1042 & \textbf{195.65} & 1,026.08 & 0/800 (0.0\%) & 1,026.08 & 67/800 (8.4\%) & $+80.9\%$ \\
2042 & \textbf{195.65} & 1,208.07 & 0/800 (0.0\%) & 1,208.07 & 79/800 (9.9\%) & $+83.8\%$ \\
\bottomrule
\end{tabular}}
\end{table}

% ----------------------------------------------------------------------------
\section{Decoded-Feasibility: Definitions}\label{app:feasibility}

The decoded-feasibility package for each baseline run reports 7 fields computed
from the final returned augmented binary state
$(\xvec, \mathbf{w}) \in \{0,1\}^{n_{\text{aug}}}$.

\paragraph{A. Native-variable feasibility.}
\begin{align*}
  \text{cardinality}     &= \textstyle\sum_{i=1}^n x_i, \\
  \text{card\_violation} &= \bigl|\textstyle\sum_{i=1}^n x_i - K\bigr|, \\
  \text{native feasible} &= [\text{card\_violation} = 0].
\end{align*}

\paragraph{B. Auxiliary consistency.}
For each Rosenberg auxiliary $w_{ij}$ intended to equal $x_i x_j$:
\begin{align*}
  \text{aux\_viol\_count} &= \#\{(i,j) : w_{ij} \neq x_i x_j\}, \\
  \text{aux\_viol\_rate}  &= \text{aux\_viol\_count} / |\mathcal{T}|, \\
  \text{FP}               &= \#\{(i,j) : w_{ij}=1 \wedge x_i x_j=0\}, \\
  \text{FN}               &= \#\{(i,j) : w_{ij}=0 \wedge x_i x_j=1\}.
\end{align*}

\paragraph{C. Objective decomposition and penalty fraction.}
\begin{equation}
  Q_{\text{aug}}(\xvec,\mathbf{w})
  = f_{\text{native}}(\xvec)
  + \underbrace{\lambda_K\!\Bigl(\textstyle\sum_i x_i - K\Bigr)^2}_{\text{card.\ pen.}}
  + \underbrace{\lambda_R \textstyle\sum_{ij} \mathrm{Ros}_{ij}(x_i,x_j,w_{ij})}_{\text{Rosenberg pen.}}
\end{equation}
\begin{equation}
  \text{penalty\_fraction}
  = \frac{|\text{card.\ pen.}| + |\text{Rosenberg pen.}|}{|Q_{\text{aug}}|}.
\end{equation}

For SA in this benchmark, the measured cardinality and Rosenberg penalty terms
are zero at all three seeds (penalty\_fraction $= 0.00\%$), yet the decoded
native objective remains far worse than HAMD.  This indicates that poor native
quality cannot be attributed solely to explicit penalty violations.

% ----------------------------------------------------------------------------
\section{Ablation Time-to-Target Curves}\label{app:ablation}

\begin{table}[h]
\centering
\caption{Full ablation TTT curves ($n=200$, $K=40$, seed 42, 60s budget).
  Numbers are the best native objective at 10/25/50/75/100\% of elapsed budget.}
\label{tab:app-ablation-ttt}
\begin{tabular}{@{}l r r r r r r@{}}
\toprule
Mode & Obj & $t_{10\%}$ & $t_{25\%}$ & $t_{50\%}$ & $t_{75\%}$ & $t_{100\%}$ \\
\midrule
HAMD-cont   & 199.86 & 214.8 & 206.0 & 204.3 & 199.9 & 199.9 \\
HAMD-proj   & 245.22 & 245.2 & 245.2 & 245.2 & 245.2 & 245.2 \\
HAMD-polish & 200.34 & 245.2 & 245.2 & 245.2 & 245.2 & 200.3 \\
HAMD-full   & \textbf{195.65} & 245.2 & 245.2 & 245.2 & 245.2 & 195.7 \\
\bottomrule
\end{tabular}
\end{table}

% ----------------------------------------------------------------------------
\section{Exact Small-Size Calibration Details}\label{app:exact}

\begin{table}[h]
\centering
\caption{Exact calibration details ($n = 20, 25, 30$; 3 seeds $\times$ 10s each).
  All 9 HAMD trials find the exact global optimum.}
\label{tab:app-exact}
\begin{tabular}{@{}r r r r r r r@{}}
\toprule
$n$ & $K$ & $|\mathcal{F}|$ & $|\mathcal{T}|$ & Exact opt & HAMD exact & Enum.\ time \\
\midrule
20 & 4 &     4,845 & 29 & $-0.178866$ & 3/3 & 0.14\,s \\
25 & 5 &    53,130 & 35 & $-0.330690$ & 3/3 & 2.37\,s \\
30 & 6 &   593,775 & 41 & $-0.435526$ & 3/3 & 28.93\,s \\
\bottomrule
\end{tabular}
\end{table}

% ----------------------------------------------------------------------------
\section{Limitations Summary}\label{app:limitations}

\begin{enumerate}[leftmargin=*]
\item \textbf{Single-seed scaling.}  Results at $n \geq 300$ use seed 42 only.
  Multi-seed campaigns at larger sizes are needed for confidence intervals.

\item \textbf{Surrogate-space baselines only.}  SA and tabu are run on the
  Rosenberg-quadratized surrogate.  We do not compare against a native-space
  discrete higher-order local search applied directly to the cubic objective
  $f(\xvec)$.  Such a baseline might close part of the observed gap and is an
  important open comparison for future work.

\item \textbf{ILS contribution.}  The ablation shows the ILS post-processing
  stage is necessary.  Claims should attribute the result to the full hybrid,
  not to continuous dynamics alone.

\item \textbf{Instance construction parameters.}  Fixed $\alpha_{\text{cubic}} = 4.0$,
  $n_{\text{sectors}} = 10$, $\lambda_R = 10.0$.  Gap magnitude may vary under
  different instance constructions.

\item \textbf{Baseline engineering.}  SA and tabu use standard single-bit-flip
  neighborhoods on the augmented QUBO.  More engineered baselines might narrow
  the gap.
\end{enumerate}

% ============================================================================
% CODE AVAILABILITY
% ============================================================================
\section*{Code Availability}\label{sec:code}

A reference implementation of the HAMD solver, benchmark instance files, and
reproduction scripts for the community-accessible experiments reported in this
paper are available at:
\begin{center}
  \url{https://github.com/symplectic-opt/hamd-community}
\end{center}
The repository includes the \texttt{NativeCubicHAMD} solver, baseline solvers
(SA and Tabu on Rosenberg-quadratized QUBO), cubic HUBO and portfolio instance
generators, and exact small-size validation (\Cref{app:exact}).
Instances at $n = 200$, $K = 40$ used in the three-seed study are included
directly.  Noncommercial use only; see the repository license
\citep{hamd_community_2026}.

% ============================================================================
% BIBLIOGRAPHY
% ============================================================================
\bibliographystyle{plain}
\bibliography{hamd_refs}

\end{document}